\documentclass{ws-p8-50x6-00}

\begin{document}

\title{On The Cosmic Origins Of Carbon And Nitrogen}

\author{R.B.C. Henry}

\address{Department of Physics \& Astronomy, University of Oklahoma, Norman, OK  73019,
USA\\E-mail: henry@mail.nhn.ou.edu}

\author{M.G. Edmunds}

\address{Department of Physics \& Astronomy, Cardiff University, P.O. Box 913, Cardiff, CF23YB Wales, UK\\
E-mail: mge@astro.cf.ac.uk}  

\author{J. K{\"o}ppen}

\address{Observatoire Astronomique, UMR. 7550, 11 rue de l'Universit{\'e}, F-67000 Strasbourg, France\\
E-mail: koppen@astro.u-strasbg.fr}  


\maketitle

\abstracts{
We employ analytical and numerical chemical evolution models to study observed trends in abundance ratios involving carbon, nitrogen, and oxygen. Several sets of published stellar yields for both intermediate-mass and massive stars are considered, and the most appropriate sets are selected through the use of analytical models. These yields are then used in the numerical models to match observed data trends in C/O, N/O, and O/H. We conclude that the principal production site for carbon is massive stars, while that for nitrogen is intermediate-mass stars.}

\section{Introduction}

Carbon and nitrogen are among the most abundant chemical elements, and of
obvious importance for life. The main nuclear processes that generate these two elements are reasonably well understood: the carbon must come from the triple-alpha reaction of helium while nitrogen originates in CNO processing by the conversion of carbon and oxygen during hydrogen burning. A lingering problem, though, has been the lack of knowledge of which {\it sites} are most important for their generation: in particular, do they come mainly from short-lived massive stars (MS) or from longer lived progenitors of asymptotic giant branch stars, i.e intermediate-mass stars (IMS)? Since necessary threshold temperatures for the production of both carbon and nitrogen are reached in both stellar types, identification of C and N production sites is difficult.

We report on the results of a study in which numerical chemical evolution models are used to match trends in nebular emission line and stellar absorption line measurements of C, N, and O, where oxygen in this case is employed as the tracer of metallicity. The basis of our study is the abundance data compiled by Henry \& Worthey\cite{hw}. Figure 1A is a plot of log(C/O) versus 12+log(O/H) by number, while Fig.~1B is the same but for log(N/O). Typical uncertainties are shown in the upper left of both figures; heavy lines are model results discussed below.

\section{Results \& Conclusions}

Our goal is to calculate numerical chemical evolution models to match trends visible in the data, at the same time employing analytical models of the data to select appropriate stellar yields for C, N, and O. Despite the scatter, we move forward by assuming a positive trend between C/O and O/H in Fig.~1A, indicating a metallicity dependence for carbon production. In Fig.~1B, we assume that the zero-slope behavior at low metallicity is commensurate with primary nitrogen production (independent of metallicity), while the clear ascending threshold for O/H above 12+log(O/H)=8.3 signifies the dominance of secondary nitrogen production (metallicity-sensitive). 

The numerical models assume a one-zone open box and include infall but not outflow in the calculations. The code comprises standard differential equations, a quadratic Schmidt law with adjustable efficiency for star formation, a Salpeter IMF, and an infall rate which was taken to be a decreasing exponential function of time with a characteristic time of 4~Gyr.

Stellar yields are the most crucial bit of model input. Their selection was based on the construction of analytical fits to the data trends by assuming simple functional forms of stellar yield dependence on metallicity. Coefficients of successful fits were then compared with coefficients describing published yields. In this way, we chose to adopt Maeder's\cite{m} yields for massive stars, in which carbon production is greatly enhanced by stellar mass loss whose rate is metallicity-sensitive. This important feature explains the secondary behavior of carbon production seen in Fig.~1A. For IMS, we adopted the yields of van~den~Hoek \& Groenewegen\cite{vdhg}, whose primary and secondary nitrogen production rates predict results which are very consistent with observed behavior of N/O.

Our model results are shown with the heavy lines in Figs.~1A,B. Models A, B, and C differ in star formation efficiency, which increases by five times each between A/B and B/C. Behavior of C/O among these models is unchanged because all C and O are produced by the same massive stars, so the abundances of these elements are unaffected by variation in star formation rate. In contrast, since N is synthesized by IMS, the evolutionary time delay of these stars compared with MS render the behavior of N/O sensitive to the temporal changes in the star formation rate. So, the greater the SFR at early times (e.g. models C vs. A), the higher O/H rises (greater rightward shift) before N/O can catch up.

Our {\it conclusions} are: (1)~Carbon is produced mostly ($>$90\%) by massive stars, and thus this element experiences no production delay relative to the synthesis of oxygen. The secondary character of its production stems from the metal-sensitive mass loss rate assumed in Maeder's massive star yields;
(2)~Nitrogen is produced mostly ($>$90\%) by intermediate-mass stars. At 12+log(O/H)$<$8.3, N synthesis is independent of metallicity, while above this value its production is sensitive to it. (3)~The characteristic delay in nitrogen production relative to that of carbon and oxygen is 250~Myr, and the relevant stellar masses for its production are 4-8M$_{\odot}$.
(4)~The low N/O and O/H values observed in low surface brightness galaxies do not necessarily imply young ages for these systems, but may instead be explained by invoking low efficiency star formation rates. A full discussion of our analysis is available in Henry, Edmunds, \& K{\"o}ppen\cite{hek}.

\begin{figure}[t]
%\figurebox{20pc}{15pc}{} % to have a box alone
\epsfxsize=14pc % will enlarge or reduce the postscript figures based on the xsize
\epsfbox{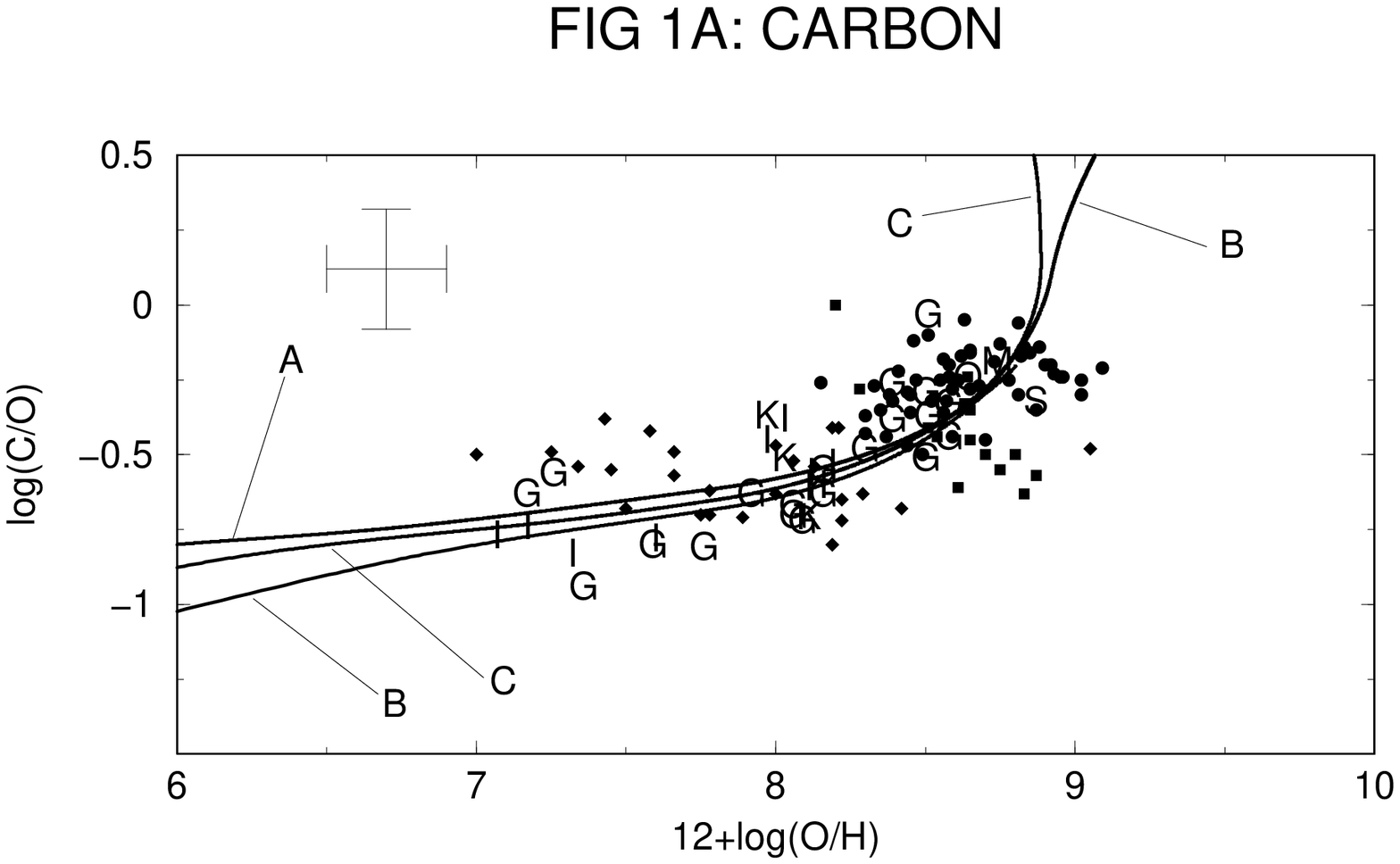} % postscript image file name
\epsfxsize=14pc % will enlarge or reduce the postscript figures based on the xsize
\epsfbox{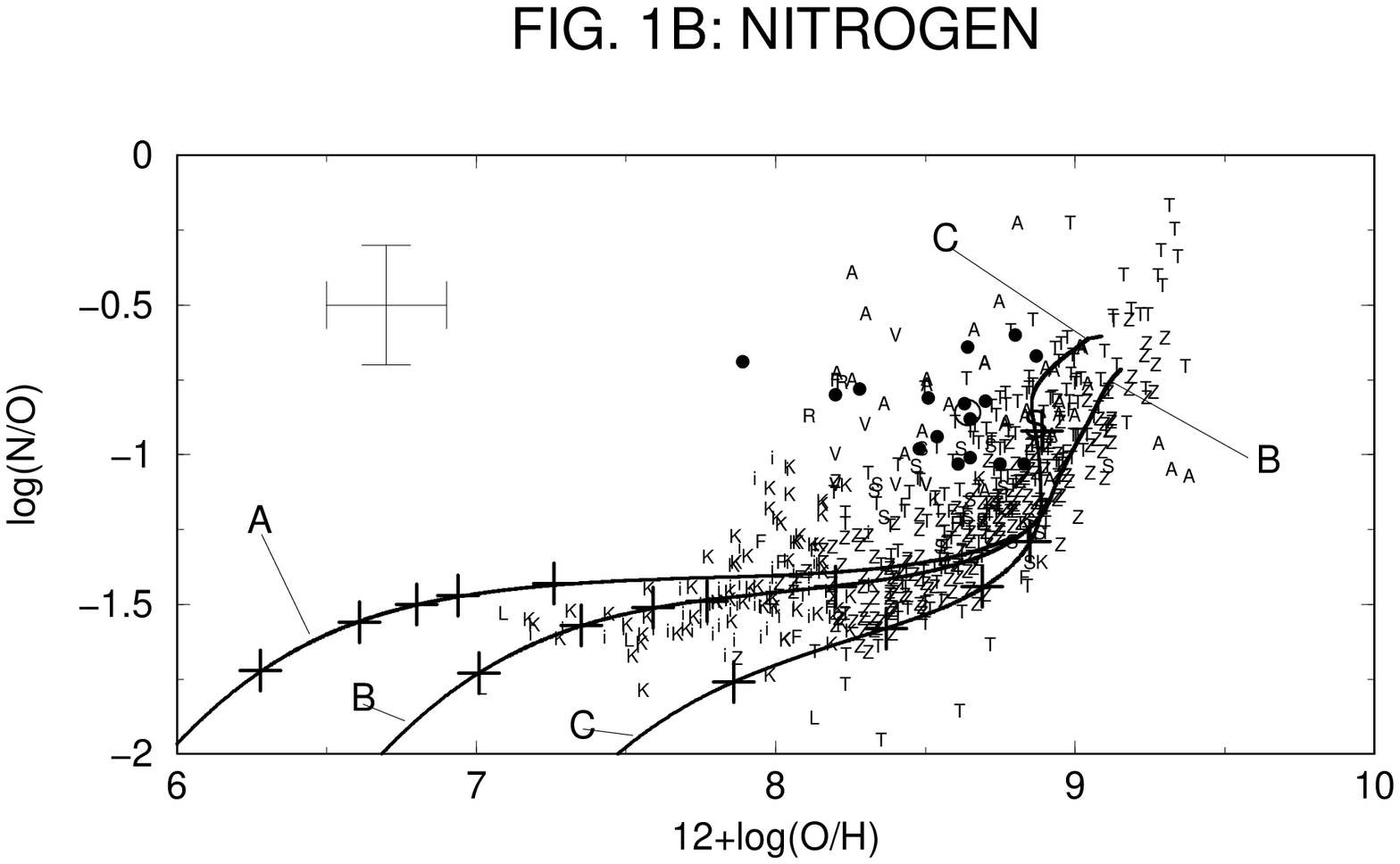} % postscript image file name
%\caption{}
\end{figure}

\section*{Acknowledgments}

RBCH gratefully acknowledges support from NSF grant AST-9819123.

\end{document}